\documentclass[review,3p,times,12pt,onecolumn]{elsarticle}
\biboptions{sort&compress}

\usepackage{amssymb}
\usepackage{lipsum}
\usepackage[T2A,T1]{fontenc}

\journal{Annals of Physics}

\begin{document}

\begin{frontmatter}

\title{Nondipole interaction between two uniformly magnetized spheres and its relation to superconducting levitation}

\author[first]{Denis Nikolaevich Sob'yanin \fontencoding{T2A}\selectfont(Денис Николаевич Собьянин)\fontencoding{T1}\selectfont}
\affiliation[first]{organization={I. E. Tamm Division of Theoretical Physics, P. N. Lebedev Physical Institute of the Russian Academy of Sciences},
            addressline={Leninskii Prospekt 53},
            city={Moscow},
            postcode={119991},
            country={Russia}}

\begin{abstract}
Analytically solving the magnetostatic Maxwell equations in the bispherical coordinates, we calculate the magnetic field around two uniformly magnetized spheres oriented so that their magnetic moments are parallel to the axis passing through the centers of the spheres. We demonstrate that, contrary to what is often claimed in the literature, the magnetic interaction between such spheres is not equivalent to the interaction between two point magnetic dipoles placed in the centers of the spheres. The nonzero levitation force acting on a uniformly magnetized sphere or a point magnetic dipole above a superconducting sphere in the ideal Meissner state is a clear manifestation of the non-equivalence.
\end{abstract}


\end{frontmatter}

\section{Introduction}

As is well known \cite{Jackson1975}, the magnetic field outside a uniformly magnetized sphere is identical to the magnetic field of a point magnetic dipole. If the sphere has a radius $R_0$ and an internal magnetic field $\mathbf{B}_0$, the external magnetic field is the same as the magnetic field of the point dipole that has the magnetic moment $\mathbf{B}_0R_0^3/2$ and is placed in the center of the sphere. This established fact was employed to simplify the consideration of a more complex case of the interaction between two uniformly magnetized spheres, which was stated to be equivalent to the interaction between two point magnetic dipoles of appropriate magnitudes and orientations placed in the centers of the spheres \cite{EdwardsEtal2017}. Due to its simplicity, the stated equivalence is now coming into use in the literature \cite{StonesEtal2019,MorinEtal2020,Ucar2021,BartzShaw2023,AmbarovEtal2023}. Meanwhile, if it takes place, we encounter the following paradox: Evidently, the above statement remains unchanged if the interaction of a uniformly magnetized sphere and a point magnetic dipole is considered. If the sphere has the zero magnetization, then, according to the statement, it should generate the external magnetic field equivalent to the field of the point dipole with the zero magnetic moment; in other words, the sphere should not generate any external magnetic field at all. We arrive at the situation where the point magnetic dipole is placed in the zero magnetic field of the nonmagnetized sphere and, therefore, does not experience any force. A possible physical realization of a nonmagnetized sphere is a superconducting sphere in the ideal Meissner state, when any internal magnetic field is expelled \cite{MeissnerOchsenfeld1933}. However, a magnetic dipole above a superconducting sphere is known to experience a nonzero levitation force \cite{Lin2006,Palaniappan2009}. In this paper, we resolve the described paradox and analytically find the magnetic field outside two uniformly magnetized spheres and the corresponding interaction force between them from the first principles. We will see that the substitution of the dipoles for the spheres, though seeming at first glance natural, leads to the results different from those following directly from the Maxwell equations.

\section{Materials and methods}

Let us consider two uniformly magnetized spheres of radii $R_1$ and $R_2$ separated by a distance~$D$, which is measured between the centers of the spheres, so that $D>R_1+R_2$. The spheres are assumed to have some uniform internal magnetic fields $\mathbf{B}_1$ and $\mathbf{B}_2$. We define the $z$ axis as the axis passing through the centers of the spheres and being directed from sphere 1 to sphere~2. In this paper we restrict ourselves to considering a purely axisymmetric case, when both the fields are parallel to the $z$ axis, which is then the axis of symmetry,
\begin{equation}
\label{B1B2parallelToZ}
\mathbf{B}_1=B_1\mathbf{e}_z,\quad\mathbf{B}_2=B_2\mathbf{e}_z,
\end{equation}
where $\mathbf{e}_z$ is the unit vector showing the direction of the increasing $z$ coordinate and $B_1$ and $B_2$ are the magnitudes of the fields. The magnitudes can be positive, zero, or negative, so that $B_i>0$ (or $B_i<0$) corresponds to the case in which sphere~$i$, where $i=1$ or~$2$, has the internal magnetic field aligned (or antialigned) with the $z$ axis, its absolute value being $|B_i|$, while $B_i=0$ corresponds to the nonmagnetized sphere~$i$.

The magnetic field $\mathbf{B}$ outside the spheres satisfies the magnetostatic Maxwell equations
\begin{equation}
\label{divBrotBzero}
\nabla\cdot\mathbf{B}=0,\quad\nabla\times\mathbf{B}=0
\end{equation}
and the boundary conditions at the surfaces of the spheres reflecting the continuity of the normal component of the magnetic field,
\begin{equation}
\label{continuousBn}
\mathbf{n}_1\cdot\mathbf{B}=\mathbf{n}_1\cdot\mathbf{B}_1,\quad\mathbf{n}_2\cdot\mathbf{B}=\mathbf{n}_2\cdot\mathbf{B}_2,
\end{equation}
where $\mathbf{n}_1$ (or $\mathbf{n}_2$) is the outward unit normal to the surface of sphere 1 (or~2) and where $\mathbf{B}$ is taken just above the surface of the corresponding sphere. Instead of the vector equations~(\ref{divBrotBzero}), after substituting
\begin{equation}
\label{Bgradpsi}
\mathbf{B}=\nabla\psi,
\end{equation}
where $\psi$ is the magnetic potential, we may deal with the Neumann problem for one scalar Laplace equation
\begin{equation}
\label{LaplaceEq}
\Delta\psi=0
\end{equation}
with the corresponding boundary conditions
\begin{equation}
\label{NeumannBC}
\frac{\partial\psi}{\partial\mathbf{n}_1}=\mathbf{n}_1\cdot\mathbf{B}_1,\quad\frac{\partial\psi}{\partial\mathbf{n}_2}=\mathbf{n}_2\cdot\mathbf{B}_2
\end{equation}
at the surfaces of the spheres.

\begin{figure*}
\centering
\includegraphics[width=13cm]{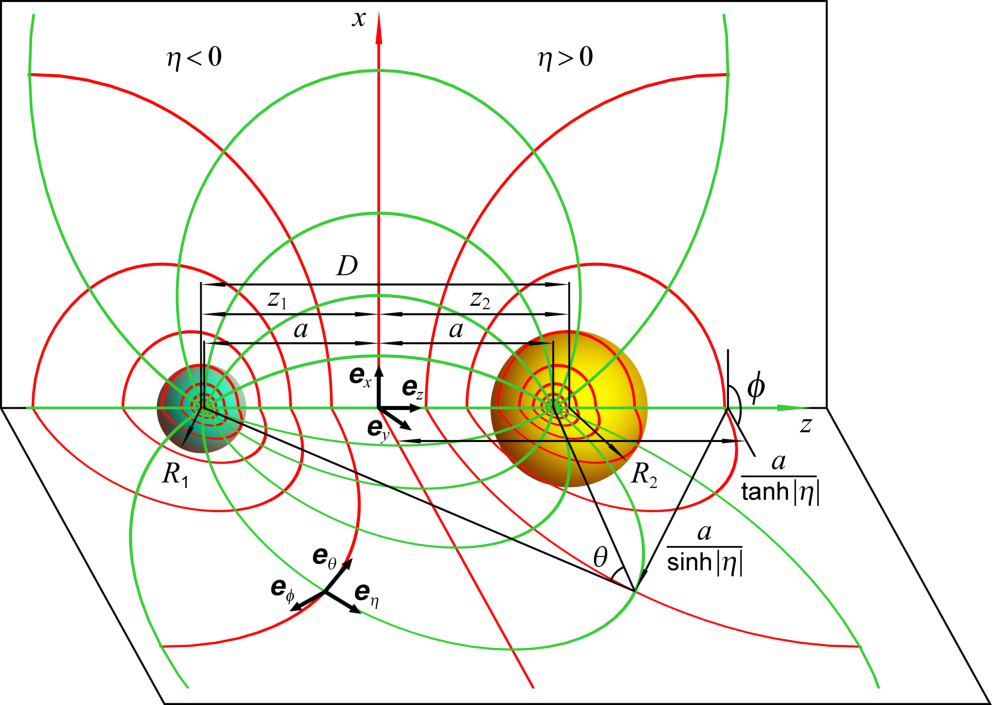}
\caption{Bispherical coordinates.}
\label{fig1}
\end{figure*}

To solve the initial problem (\ref{divBrotBzero}) and (\ref{continuousBn}) or the equivalent problem (\ref{LaplaceEq}) and~(\ref{NeumannBC}), we employ the bispherical (or three-dimensional bipolar) coordinates (Fig.~\ref{fig1}), which are orthogonal, have spheres as coordinate surfaces, and allow the separation of variables for the Laplace equation after introducing a separating multiplier \cite{Jeffery1912,MorseFeshbach1953,Lebedev1965,MoonSpencer1988}. Due to these useful properties the bispherical coordinates, together with the methods of images and inversion transformation \cite{Weiss1944,LinJin1990,Palaniappan2007,GaoEtal2012}, are often used in various electromagnetic and hydrodynamic problems involving two spheres \cite{Endo1938,Love1975,WarrenCuthrell1975,Stoy1989a,Stoy1989b,ChaumetDufour1998,Lekner2011,VafeasEtal2015,GilbertGiacomin2019,Majic2021,Lekner2021}. Two positive quantities
\begin{equation}
\label{z12}
z_1=\frac{D^2+R_1^2-R_2^2}{2D},\quad z_2=\frac{D^2-R_1^2+R_2^2}{2D}
\end{equation}
with the property $z_1+z_2=D$ divide the distance between the centers of the spheres into two parts. We adopt the dividing point as the origin of the usual Cartesian coordinates with the $z$ axis defined above and additionally define the $x$ axis with the corresponding unit vector $\mathbf{e}_x$ as being orthogonal to the $z$ axis and defining the zero half-plane, which consists of points with nonnegative $x$ coordinates, and the $y$ axis with the corresponding unit vector $\mathbf{e}_y$ as being obtained by rotation of the $x$ axis about the $z$ axis through angle $\pi/2$. In these Cartesian coordinates the center of sphere 1 has coordinates $(0,0,-z_1)$ and the center of sphere 2 has coordinates $(0,0,z_2)$. The bispherical coordinates $(\eta,\theta,\phi)$ are introduced via
\begin{equation}
\label{bisphericalCoordinates}
x=\frac{a\sin\theta\cos\phi}{\cosh\eta-\cos\theta},
\quad
y=\frac{a\sin\theta\sin\phi}{\cosh\eta-\cos\theta},
\quad
z=\frac{a\sinh\eta}{\cosh\eta-\cos\theta},
\end{equation}
where
\begin{equation}
\label{a}
a=\frac{1}{2D}\sqrt{(D+R_1+R_2)(D+R_1-R_2)(D-R_1+R_2)(D-R_1-R_2)}
\end{equation}
is the semifocal distance.

The two foci have Cartesian coordinates $(0,0,-a)$ and $(0,0,a)$ and lie inside spheres 1 and 2, respectively, not coinciding with their centers. The coordinate $\phi$ ranges from $0$ to $2\pi$ and corresponds to the standard azimuthal coordinate in the cylindrical or spherical coordinates with the $z$ axis taken as the polar axis, so that the coordinate surface is the half-plane obtained from the zero half-plane by its rotation about the $z$ axis through angle~$\phi$. The coordinate $\theta$ ranges from $0$ to $\pi$ and corresponds to the angle between the two vectors directed from a given point to the two foci of the bispherical coordinates, so that the coordinate surface is obtained by revolution about the $z$ axis of the circular arc which has the foci as the endpoints and at every point of which the angle subtended by the interval between the foci is~$\theta$. The coordinate $\eta$ ranges from $-\infty$ to $+\infty$ and corresponds to spherical coordinate surfaces having radii $R=a/\sinh|\eta|$ and being centered at points $(0,0,a/\tanh\eta)$ when $\eta\neq0$. The coordinate $\eta=0$ corresponds to the plane being orthogonal to the $z$ axis and passing through the origin, thus containing the $x$ and $y$ axes, $\eta=-\infty$ to the focus $(0,0,-a)$, and $\eta=+\infty$ to the focus $(0,0,a)$.The two positive quantities $\eta_1$ and $\eta_2$ defined via
\begin{equation}
\label{etas}
\cosh\eta_1=\frac{z_1}{R_1},\quad \cosh\eta_2=\frac{z_2}{R_2}
\end{equation}
correspond to the surfaces of spheres 1 and~2, respectively, and the range
\begin{equation}
\label{etaRange}
-\eta_1<\eta<\eta_2
\end{equation}
corresponds to the space outside the spheres, in which we wish to determine the magnetic field $\mathbf{B}$. The orthogonal unit vectors
\begin{equation}
\label{eEta}
\mathbf{e}_\eta=-\frac{\sinh\eta\sin\theta}{\cosh\eta-\cos\theta}(\cos\phi\,\mathbf{e}_x+\sin\phi\,\mathbf{e}_y)-\frac{\cosh\eta\cos\theta-1}{\cosh\eta-\cos\theta}\,\mathbf{e}_z,
\end{equation}
\begin{equation}
\label{eTheta}
\mathbf{e}_\theta=\frac{\cosh\eta\cos\theta-1}{\cosh\eta-\cos\theta}(\cos\phi\,\mathbf{e}_x+\sin\phi\,\mathbf{e}_y)-\frac{\sinh\eta\sin\theta}{\cosh\eta-\cos\theta}\,\mathbf{e}_z,
\end{equation}
\begin{equation}
\label{ePhi}
\mathbf{e}_\phi=-\sin\phi\,\mathbf{e}_x+\cos\phi\,\mathbf{e}_y,
\end{equation}
which are obtained from (\ref{bisphericalCoordinates}) by differentiating and dividing by the Lam\'{e} coefficients, can unambiguously be constructed at every point except the foci and correspond to the directions of the increasing respective coordinates $(\eta,\theta,\phi)$. Particularly, the vector $\mathbf{e}_\eta$ is orthogonal to the spherical coordinate surfaces determined by the condition $\eta=const$, so that at the surfaces of spheres 1 and 2 we have the following relations of $\mathbf{e}_\eta$ with the outward normals:
\begin{equation}
\label{etaNormal}
\mathbf{e}_\eta|_{\eta=-\eta_1}=\mathbf{n}_1,
\quad
\mathbf{e}_\eta|_{\eta=\eta_2}=-\mathbf{n}_2.
\end{equation}

Since we deal with the axisymmetric situation, when the internal magnetic fields of the two uniformly magnetized spheres are parallel to the $z$ axis, the boundary conditions at the surfaces of the spheres are axisymmetric; therefore, the magnetic potential $\psi=\psi(\eta,\theta)$ is axisymmetric and independent of the azimuthal coordinate~$\phi$. The external magnetic field is also axisymmetric,
\begin{equation}
\label{axisymmetricB}
\mathbf{B}=B_\eta\mathbf{e}_\eta+B_\theta\mathbf{e}_\theta,
\end{equation}
and does not contain the azimuthal component parallel to $\mathbf{e}_\phi$, so that $B_\phi=0$, while the remaining two components $B_\eta=B_\eta(\eta,\theta)$ and $B_\theta=B_\theta(\eta,\theta)$ are independent of~$\phi$ and, as follows from~(\ref{Bgradpsi}), are expressed via the magnetic potential as
\begin{equation}
\label{BetaBtheta}
B_\eta=\frac{\cosh\eta-\cos\theta}{a}\frac{\partial\psi}{\partial\eta},
\quad
B_\theta=\frac{\cosh\eta-\cos\theta}{a}\frac{\partial\psi}{\partial\theta}.
\end{equation}
The independent axisymmetric harmonics satisfying the Laplace equation (\ref{LaplaceEq}) have the form
\begin{equation}
\label{axisymmetricHarmonic+}
\psi_l^+(\eta,\theta)=\sqrt{2(\cosh\eta-\cos\theta)}\,\mathrm{P}_l(\cos\theta)e^{(l+1/2)\eta},
\end{equation}
\begin{equation}
\label{axisymmetricHarmonic-}
\psi_l^-(\eta,\theta)=\sqrt{2(\cosh\eta-\cos\theta)}\,\mathrm{P}_l(\cos\theta)e^{-(l+1/2)\eta},
\end{equation}
where $\sqrt{2(\cosh\eta-\cos\theta)}$ is the aforementioned separating multiplier, $\mathrm{P}_l(\cos\theta)$ is the Legendre polynomial, and $l$ is a nonnegative integer; therefore, the general axisymmetric magnetic potential is
\begin{equation}
\label{generalAxisymmetricSolution}
\psi=\sqrt{2(\cosh\eta-\cos\theta)}\sum_{l=0}^\infty \mathrm{P}_l(\cos\theta) \bigl(A_l^+ e^{(l+1/2)\eta} + A_l^- e^{-(l+1/2)\eta}\bigr),
\end{equation}
where $A_l^+$ and $A_l^-$ are arbitrary constants.

\section{Results}

To find the components of the magnetic field (\ref{axisymmetricB}) outside the two uniformly magnetized spheres, we substitute the general expression for the magnetic potential into~(\ref{BetaBtheta}). Differentiating (\ref{generalAxisymmetricSolution}) with respect to $\eta$ and~$\theta$, multiplying by the separating multiplier $\sqrt{2(\cosh\eta-\cos\theta)}$, using the contiguous relations
\begin{equation}
(2l+1)\mu \mathrm{P}_l(\mu)=(l+1)\mathrm{P}_{l+1}(\mu)+l\mathrm{P}_{l-1}(\mu),
\end{equation}
\begin{equation}
(2l+1)\mu\mathrm{P}^1_l(\mu)=l\mathrm{P}^1_{l+1}(\mu)+(l+1)\mathrm{P}^1_{l-1}(\mu),
\end{equation}
\begin{equation}
(2l+1)\sqrt{1-\mu^2}\,\mathrm{P}_l(\mu)=\mathrm{P}^1_{l-1}(\mu)-\mathrm{P}^1_{l+1}(\mu)
\end{equation}
for the Legendre polynomials $\mathrm{P}_l(\mu)$ and the associated Legendre functions \cite{GradshteynRyzhik2015}
\begin{equation}
\mathrm{P}^1_l(\mu)=-\sqrt{1-\mu^2}\frac{d\mathrm{P}_l(\mu)}{d\mu},\quad \mu=\cos\theta,
\end{equation}
regrouping the resulting terms, and multiplying by $\sqrt{2(\cosh\eta-\cos\theta)}/2a$, we eventually obtain
\begin{equation}
B_\eta = \frac12 \sqrt{2(\cosh\eta-\cos\theta)}\sum_{l=0}^\infty \mathrm{P}_l(\cos\theta) \bigl[l\bigl(\alpha_l^+ e^{(l-1/2)\eta} - \alpha_l^- e^{-(l-1/2)\eta}\bigr)
- (l+1)\bigl(\alpha_{l+1}^+ e^{(l+3/2)\eta} - \alpha_{l+1}^- e^{-(l+3/2)\eta}\bigr)\bigr],
\label{Beta}
\end{equation}
\begin{equation}
B_\theta = \frac12 \sqrt{2(\cosh\eta-\cos\theta)}\sum_{l=1}^\infty \mathrm{P}^1_l(\cos\theta) \bigl(\alpha_l^+ e^{(l-1/2)\eta} + \alpha_l^- e^{-(l-1/2)\eta}
- \alpha_{l+1}^+ e^{(l+3/2)\eta} - \alpha_{l+1}^- e^{-(l+3/2)\eta}\bigr),
\label{Btheta}
\end{equation}
where $\alpha_l^+=(A_l^+ - A_{l-1}^+)/a$ and $\alpha_l^-=(A_l^- - A_{l-1}^-)/a$ are arbitrary constants. Thus, the problem of finding the magnetic field is reduced to the problem of choosing the constants $\alpha_l^+$ and $\alpha_l^-$ so as to satisfy the boundary conditions~(\ref{continuousBn}).

Noting that $\mathbf{e}_z=\nabla z$, we express the right-hand sides of (\ref{NeumannBC}) via the normal derivatives analogously to the left-hand sides,
\begin{equation}
\frac{\partial\psi}{\partial\mathbf{n}_1}=B_1\frac{\partial z}{\partial\mathbf{n}_1},
\quad
\frac{\partial\psi}{\partial\mathbf{n}_2}=B_2\frac{\partial z}{\partial\mathbf{n}_2}.
\end{equation}
It is evident from (\ref{Bgradpsi}) and (\ref{etaNormal})--(\ref{BetaBtheta}) that the normal derivative at the surface of sphere 1 or 2 is proportional to the partial derivative with respect to~$\eta$; the boundary conditions can then be rewritten as
\begin{equation}
\label{axisymmetricBC}
\frac{\partial\psi_1}{\partial\eta}\Bigr|_{\eta=-\eta_1}=\frac{\partial\psi_2}{\partial\eta}\Bigr|_{\eta=\eta_2}=0,
\end{equation}
where the auxiliary potentials
\begin{equation}
\label{auxiliaryPotentials}
\psi_1=\psi-B_1 z,
\quad
\psi_2=\psi-B_2 z
\end{equation}
are introduced. Since these potentials will only be used for finding the constants that provide the fulfillment of the boundary conditions at the surfaces of the magnetized spheres, the behavior of the potentials near the respective surfaces is only important, and it is sufficient to consider the auxiliary potential $\psi_1$ for negative~$\eta$, which corresponds to the half-space $z<0$, where sphere 1 lies, and the auxiliary potential $\psi_2$ for positive~$\eta$, which corresponds to the half-space $z>0$, where sphere 2 lies. The representation of the coordinate $z$ via the series analogous to (\ref{generalAxisymmetricSolution}) has the form \cite{Lekner2021}
\begin{equation}
\label{zDecomposition}
z=a\,\mathrm{sgn}\,\eta\,\sqrt{2(\cosh\eta-\cos\theta)}\sum_{l=0}^\infty (2l+1)\mathrm{P}_l(\cos\theta)e^{-(l+1/2)|\eta|},
\end{equation}
and the auxiliary potentials become
\begin{equation}
\psi_1=\sqrt{2(\cosh\eta-\cos\theta)}\sum_{l=0}^\infty \mathrm{P}_l(\cos\theta) \bigl\{\bigl[A_l^+ +(2l+1)B_1\bigr]e^{(l+1/2)\eta} + A_l^- e^{-(l+1/2)\eta}\bigr\}
\label{psi1}
\end{equation}
for $\eta<0$ and
\begin{equation}
\psi_2=\sqrt{2(\cosh\eta-\cos\theta)}\sum_{l=0}^\infty \mathrm{P}_l(\cos\theta) \bigl\{A_l^+ e^{(l+1/2)\eta} + \bigl[A_l^- -(2l+1)B_2\bigr]e^{-(l+1/2)\eta}\bigr\}
\label{psi2}
\end{equation}
for $\eta>0$. By analogy with (\ref{Beta}), we obtain
\begin{eqnarray}
\sqrt{2(\cosh\eta-\cos\theta)}\,\frac{\partial\psi_1}{\partial\eta}=\sum_{l=0}^\infty \mathrm{P}_l(\cos\theta) \bigl\{l\bigl[(\alpha_l^+  + 2B_1) e^{(l-1/2)\eta} - \alpha_l^- e^{-(l-1/2)\eta}\bigr]
\nonumber\\
-(l+1)\bigl[(\alpha_{l+1}^+ + 2B_1) e^{(l+3/2)\eta} - \alpha_{l+1}^- e^{-(l+3/2)\eta}\bigr]\bigr\},
\label{sqrtdpsi1deta}
\end{eqnarray}
\begin{eqnarray}
\sqrt{2(\cosh\eta-\cos\theta)}\,\frac{\partial\psi_2}{\partial\eta}=\sum_{l=0}^\infty \mathrm{P}_l(\cos\theta) \bigl\{l\bigl[\alpha_l^+ e^{(l-1/2)\eta} - (\alpha_l^- - 2B_2) e^{-(l-1/2)\eta}\bigr]
\nonumber\\
-(l+1)\bigl[\alpha_{l+1}^+ e^{(l+3/2)\eta} - (\alpha_{l+1}^- - 2B_2) e^{-(l+3/2)\eta}\bigr]\bigr\}.
\label{sqrtdpsi2deta}
\end{eqnarray}
From (\ref{axisymmetricBC}), (\ref{sqrtdpsi1deta}), and (\ref{sqrtdpsi2deta}) we arrive at the following infinite system of equations:
\begin{equation}
l\bigl[\alpha_l^- e^{(l-1/2)\eta_1} - (\alpha_l^+  + 2B_1) e^{-(l-1/2)\eta_1}\bigr]
= (l+1)\bigl[\alpha_{l+1}^- e^{(l+3/2)\eta_1} - (\alpha_{l+1}^+ + 2B_1) e^{-(l+3/2)\eta_1}\bigr],
\label{alphaSE1}
\end{equation}
\begin{equation}
l\bigl[\alpha_l^+ e^{(l-1/2)\eta_2} - (\alpha_l^- - 2B_2) e^{-(l-1/2)\eta_2}\bigr]
= (l+1)\bigl[\alpha_{l+1}^+ e^{(l+3/2)\eta_2} - (\alpha_{l+1}^- - 2B_2) e^{-(l+3/2)\eta_2}\bigr].
\label{alphaSE2}
\end{equation}

To solve the system and find the constants $\alpha_l^+$ and~$\alpha_l^-$, we introduce the generating functions
\begin{equation}
\label{genFunc}
F^+(\lambda)=\sum_{l=1}^\infty l \alpha_l^+ \lambda^l,
\quad
F^-(\lambda)=\sum_{l=1}^\infty l \alpha_l^- \lambda^l.
\end{equation}
Multiplying the left- and right-hand sides of (\ref{alphaSE1}) and (\ref{alphaSE2}) by $\lambda^l$ and summing over $l$ from $0$ to~$\infty$, we obtain
\begin{equation}
\biggl(\frac{e^{\eta_1/2}}\lambda - e^{-\eta_1/2}\biggr)F^-(e^{\eta_1}\lambda)
+ \biggl(e^{\eta_1/2} - \frac{e^{-\eta_1/2}}\lambda\biggr)\biggl[F^+(e^{-\eta_1}\lambda) + 2B_1 \frac{e^{-\eta_1} \lambda}{(1 - e^{-\eta_1} \lambda)^2}\biggr] = 0,
\label{FSE1}
\end{equation}
\begin{equation}
\biggl(\frac{e^{\eta_2/2}}\lambda - e^{-\eta_2/2}\biggr)F^+(e^{\eta_2}\lambda)
+ \biggl(e^{\eta_2/2} - \frac{e^{-\eta_2/2}}\lambda\biggr)\biggl[F^-(e^{-\eta_2}\lambda) - 2B_2 \frac{e^{-\eta_2} \lambda}{(1 - e^{-\eta_2} \lambda)^2}\biggr] = 0.
\label{FSE2}
\end{equation}
Substituting $\lambda\rightarrow e^{-\eta_2}\lambda$ (or $\lambda\rightarrow e^{\eta_2}\lambda$) into (\ref{FSE1}) and $\lambda\rightarrow e^{\eta_1}\lambda$ (or $\lambda\rightarrow e^{-\eta_1}\lambda$) into (\ref{FSE2}) and combining the resulting equations so that the terms containing $F^-$ (or~$F^+$) cancel out, we may obtain an equation containing only the generating function $F^+$ (or~$F^-$). We eventually have
\begin{equation}
e^{\eta_1+\eta_2}\Phi^+(e^{\eta_1+\eta_2}\lambda) - \Phi^+(e^{-\eta_1-\eta_2}\lambda)
= 2 a e^{-\eta_1-\eta_2}\lambda \biggl[\frac{B_1}{(1 - e^{-\eta_1-\eta_2}\lambda)^3} - \frac{B_2}{(e^{-\eta_1} - e^{-\eta_2}\lambda)^3}\biggr],
\label{Phi+}
\end{equation}
\begin{equation}
e^{\eta_1+\eta_2}\Phi^-(e^{\eta_1+\eta_2}\lambda) - \Phi^-(e^{-\eta_1-\eta_2}\lambda)
= 2 a e^{-\eta_1-\eta_2}\lambda \biggl[\frac{B_1}{(e^{-\eta_2} - e^{-\eta_1}\lambda)^3} - \frac{B_2}{(1 - e^{-\eta_1-\eta_2}\lambda)^3}\biggr],
\label{Phi-}
\end{equation}
where
\begin{equation}
\label{PhiDef}
\Phi^+(\lambda)=\sum_{l=1}^\infty \beta_l^+ \lambda^l=\frac{F^+(\lambda)}{1-\lambda},
\quad
\Phi^-(\lambda)=\sum_{l=1}^\infty \beta_l^- \lambda^l=\frac{F^-(\lambda)}{1-\lambda}
\end{equation}
are new generating functions, which are defined via the previous generating functions $F^+(\lambda)$ and $F^-(\lambda)$. Substituting the associated power series into (\ref{Phi+}) and~(\ref{Phi-}), expanding the right-hand sides into power series using the relation
\begin{equation}
\frac{2\zeta}{(1-\zeta)^3}=\sum_{l=1}^\infty l(l+1)\zeta^l,
\end{equation}
and equating the coefficients of equal powers of $\lambda$ on both sides, we obtain
\begin{equation}
\label{beta12}
\beta_l^+ = l(l+1) \frac{B_1 - B_2 e^{(2l+1)\eta_1}}{e^{(2l+1)(\eta_1 + \eta_2)} - 1},
\quad
\beta_l^- = l(l+1) \frac{B_1 e^{(2l+1)\eta_2} - B_2}{e^{(2l+1)(\eta_1 + \eta_2)} - 1}.
\end{equation}
It follows from (\ref{PhiDef}) that
\begin{equation}
\label{alphaViaBeta}
l\alpha_l^+=\beta_l^+ - \beta_{l-1}^+,\quad l\alpha_l^-=\beta_l^- - \beta_{l-1}^-;
\end{equation}
therefore,
\begin{equation}
\alpha_l^+ = (l+1)\frac{B_1 - B_2 e^{(2l+1)\eta_1}}{e^{(2l+1)(\eta_1 + \eta_2)}  - 1} - (l-1)\frac{B_1 - B_2 e^{(2l-1)\eta_1}}{e^{(2l-1)(\eta_1 + \eta_2)} - 1},
\label{alpha+}
\end{equation}
\begin{equation}
\alpha_l^- = (l+1)\frac{B_1 e^{(2l+1)\eta_2} - B_2}{e^{(2l+1)(\eta_1 + \eta_2)} - 1} - (l-1)\frac{B_1 e^{(2l-1)\eta_2} - B_2}{e^{(2l-1)(\eta_1 + \eta_2)} - 1}.
\label{alpha-}
\end{equation}
The validity of (\ref{alpha+}) and (\ref{alpha-}) is verified by direct substitution into (\ref{alphaSE1}) and~(\ref{alphaSE2}).

Now let us verify that the series in (\ref{Beta}) and (\ref{Btheta}) are not formal and converge. The constants $\alpha_l^+$ and $\alpha_l^-$ can be bounded thus:
\begin{eqnarray}
|\alpha_l^+|&\leq& (l+1)\frac{|B_1| + |B_2| e^{(2l+1)\eta_1}}{e^{(2l+1)(\eta_1 + \eta_2)}  - 1} + (l-1)\frac{|B_1| + |B_2| e^{(2l-1)\eta_1}}{e^{(2l-1)(\eta_1 + \eta_2)} - 1}
\nonumber\\
&=& (l+1) e^{-(2l+1)\eta_2} \frac{|B_1| e^{-(2l+1)\eta_1} + |B_2|}{1 - e^{-(2l+1)(\eta_1 + \eta_2)}} + (l-1) e^{-(2l-1)\eta_2} \frac{|B_1| e^{-(2l-1)\eta_1} + |B_2|}{1 - e^{-(2l-1)(\eta_1 + \eta_2)}}
\nonumber\\
&\leq& (l+1) e^{-(2l-1)\eta_2} \frac{|B_1| e^{-3\eta_1} + |B_2|}{1 - e^{-3(\eta_1 + \eta_2)}} + (l-1) e^{-(2l-1)\eta_2} \frac{|B_1| e^{-3\eta_1} + |B_2|}{1 - e^{-3(\eta_1 + \eta_2)}}
\nonumber\\
&\leq& 2l e^{-(2l-1)\eta_2} \frac{|B_1| + |B_2|}{1 - e^{-3(\eta_1 + \eta_2)}}
\label{alpha+Bound}
\end{eqnarray}
and, analogously,
\begin{equation}
\label{alpha-Bound}
|\alpha_l^-| \leq 2l e^{-(2l-1)\eta_1} \frac{|B_1| + |B_2|}{1 - e^{-3(\eta_1 + \eta_2)}},
\end{equation}
where the inequalities $|b+c|\leq|b|+|c|$ and $l\geq1$ are used. It follows from (\ref{etaRange}), (\ref{alpha+Bound}), and (\ref{alpha-Bound}) that
\begin{equation}
\label{alpha+Exp}
|\alpha_l^+| e^{(l-1/2)\eta} \leq 2l e^{-(l-1/2)\eta_2} \frac{|B_1| + |B_2|}{1 - e^{-3(\eta_1 + \eta_2)}},
\end{equation}
\begin{equation}
\label{alpha+Exp1}
|\alpha_{l+1}^+| e^{(l+3/2)\eta} \leq 2(l+1) e^{-(l-1/2)\eta_2} \frac{|B_1| + |B_2|}{1 - e^{-3(\eta_1 + \eta_2)}},
\end{equation}
\begin{equation}
\label{alpha-Exp}
|\alpha_l^-| e^{-(l-1/2)\eta} \leq 2l e^{-(l-1/2)\eta_1} \frac{|B_1| + |B_2|}{1 - e^{-3(\eta_1 + \eta_2)}},
\end{equation}
\begin{equation}
\label{alpha-Exp1}
|\alpha_{l+1}^-| e^{-(l+3/2)\eta} \leq 2(l+1) e^{-(l-1/2)\eta_1} \frac{|B_1| + |B_2|}{1 - e^{-3(\eta_1 + \eta_2)}};
\end{equation}
therefore, the coefficients of the Legendre polynomials in the series in (\ref{Beta}) and (\ref{Btheta}) can be bounded thus:
\begin{eqnarray}
&&\bigl|l\bigl(\alpha_l^+ e^{(l-1/2)\eta} - \alpha_l^- e^{-(l-1/2)\eta}\bigr) - (l+1)\bigl(\alpha_{l+1}^+ e^{(l+3/2)\eta} - \alpha_{l+1}^- e^{-(l+3/2)\eta}\bigr)\bigr|
\nonumber\\
&\leq& l\bigl(|\alpha_l^+| e^{(l-1/2)\eta} + |\alpha_l^-| e^{-(l-1/2)\eta}\bigr) + (l+1)\bigl(|\alpha_{l+1}^+| e^{(l+3/2)\eta} + |\alpha_{l+1}^-| e^{-(l+3/2)\eta}\bigr)
\nonumber\\
&\leq& 2\bigl[l^2 + (l+1)^2\bigr]\bigl(e^{-(l-1/2)\eta_1} + e^{-(l-1/2)\eta_2}\bigr) \frac{|B_1| + |B_2|}{1 - e^{-3(\eta_1 + \eta_2)}}
\nonumber\\
&\leq& 4(2l+1)^2 e^{-(l-1/2)\min(\eta_1,\eta_2)} \frac{|B_1| + |B_2|}{1 - e^{-3(\eta_1 + \eta_2)}},
\label{PcoefBound}
\end{eqnarray}
\begin{eqnarray}
&&\bigl|\alpha_l^+ e^{(l-1/2)\eta} + \alpha_l^- e^{-(l-1/2)\eta} - \alpha_{l+1}^+ e^{(l+3/2)\eta} - \alpha_{l+1}^- e^{-(l+3/2)\eta}\bigr|
\nonumber\\
&\leq& |\alpha_l^+| e^{(l-1/2)\eta} + |\alpha_l^-| e^{-(l-1/2)\eta} + |\alpha_{l+1}^+| e^{(l+3/2)\eta} + |\alpha_{l+1}^-| e^{-(l+3/2)\eta}
\nonumber\\
&\leq& 2(2l+1) \bigl(e^{-(l-1/2)\eta_1} + e^{-(l-1/2)\eta_2}\bigr) \frac{|B_1| + |B_2|}{1 - e^{-3(\eta_1 + \eta_2)}}
\nonumber\\
&\leq& 4(2l+1) e^{-(l-1/2)\min(\eta_1,\eta_2)} \frac{|B_1| + |B_2|}{1 - e^{-3(\eta_1 + \eta_2)}}.
\label{P1coefBound}
\end{eqnarray}
Note that the inequalities (\ref{PcoefBound}) and (\ref{P1coefBound}) are derived for $l\geq1$, and we also need to bound the coefficient of $\mathrm{P}_l(\cos\theta)$ in (\ref{Beta}) for $l=0$:
\begin{eqnarray}
&&\bigl| -\bigl(\alpha_1^+ e^{3\eta/2} - \alpha_1^- e^{-3\eta/2}\bigr)\bigr|
= 2\frac{\bigl|B_1 \bigl(e^{3\eta_2 - 3\eta/2} - e^{3\eta/2}\bigr) + B_2 \bigl(e^{3\eta_1 + 3\eta/2} - e^{-3\eta/2}\bigr)\bigr|}{e^{3(\eta_1 + \eta_2)} - 1}
\nonumber\\
&\leq& 2\frac{|B_1|\bigl|e^{3\eta_2 - 3\eta/2} - e^{3\eta/2}\bigr| + |B_2|\bigl|e^{3\eta_1 + 3\eta/2} - e^{-3\eta/2}\bigr|}{e^{3(\eta_1 + \eta_2)} - 1}
\leq 2\bigl(|B_1|e^{-3\eta_1/2}+|B_2|e^{-3\eta_2/2}\bigr).
\label{PcoefBoundForlZero}
\end{eqnarray}

Since $|\mathrm{P}_l(\cos\theta)|\leq1$ \cite{GradshteynRyzhik2015}, we have
\begin{equation}
\bigl|\mathrm{P}_l(\cos\theta) \bigl[l\bigl(\alpha_l^+ e^{(l-1/2)\eta} - \alpha_l^- e^{-(l-1/2)\eta}\bigr)
- (l+1)\bigl(\alpha_{l+1}^+ e^{(l+3/2)\eta} - \alpha_{l+1}^- e^{-(l+3/2)\eta}\bigr)\bigr]\bigr| \leq M_l,
\label{boundBeta}
\end{equation}
where
\begin{equation}
\label{Ml0}
M_0 = 2\bigl(|B_1|e^{-3\eta_1/2}+|B_2|e^{-3\eta_2/2}\bigr),
\end{equation}
\begin{equation}
\label{Ml1}
M_{l\geq1} = 4(2l+1)^2 e^{-(l-1/2)\min(\eta_1,\eta_2)}\frac{|B_1| + |B_2|}{1 - e^{-3(\eta_1 + \eta_2)}}.
\end{equation}
Correspondingly, since \cite{Fryant1986,Lohofer1998}
\begin{equation}
|\mathrm{P}_l^m(\cos\theta)|\leq\sqrt{\frac{(l+m)!}{2(l-m)!}},\quad 1\leq|m|\leq l,
\end{equation}
we have
\begin{equation}
\label{P1Bound}
|\mathrm{P}_l^1(\cos\theta)|\leq\sqrt{\frac{l(l+1)}2}=\sqrt{\frac{(l+1/2)^2 - 1/4}2}<\frac{2l+1}{2\sqrt{2}}
\end{equation}
and, taking account of (\ref{P1coefBound}),
\begin{equation}
\bigl|\mathrm{P}^1_l(\cos\theta) \bigl(\alpha_l^+ e^{(l-1/2)\eta} + \alpha_l^- e^{-(l-1/2)\eta} - \alpha_{l+1}^+ e^{(l+3/2)\eta} - \alpha_{l+1}^- e^{-(l+3/2)\eta}\bigr)\bigr| \leq \frac{M_l}{2\sqrt{2}}.
\label{boundBtheta}
\end{equation}
Thus, the absolute value of the $l$th element of the series determining $B_\eta$ (or~$B_\theta$) is bounded by the $l$th element of the series $M_\eta$ (or~$M_\theta$), where
\begin{equation}
\label{boundingSeries}
M_\eta=\sum_{l=0}^\infty M_l,\quad M_\theta=\sum_{l=1}^\infty \frac{M_l}{2\sqrt{2}}.
\end{equation}

The convergence of the series
\begin{equation}
\sum_{l=1}^\infty M_l = 4\sqrt{p}\,\biggl(\frac{1}{1-p} + \frac{8}{(1-p)^3}\biggr) \frac{|B_1| + |B_2|}{1 - e^{-3(\eta_1 + \eta_2)}} < +\infty,
\quad
p = e^{-\min(\eta_1,\eta_2)}<1,
\label{convSeries}
\end{equation}
entails the convergence of the series $M_\eta$ and~$M_\theta$; therefore, the series in (\ref{Beta}) and (\ref{Btheta}) have passed the Weierstrass M-test \cite{Nikolsky1977,Apostol1981} and hence converge absolutely and uniformly outside the spheres, i.e., on the entire domain~(\ref{etaRange}).

It is interesting to estimate the speed of convergence. When $D\gg R_1=R_2$ and $B_1=B_2$, we have $M_l\sim32 l^2 (R_1/D)^{l-1/2}|B_1|$ for $l\gg1$. If only the terms with numbers $l\leq l_0\gg1$ are retained and the terms with numbers $l>l_0$ are dropped when the magnetic field is being calculated, then the contribution of the latter terms to $B_\eta$ cannot exceed
\begin{equation}
\label{resBeta}
\delta B_\eta\sim\frac{1}{2}\biggl(\frac{D}{R_1}\biggr)^{1/2}\sum_{l=l_0+1}^\infty M_l \sim 16l_0^2 \biggl(\frac{R_1}{D}\biggr)^{l_0}|B_1|.
\end{equation}
Therefore, the number of terms providing the relative error $\epsilon=\delta B_\eta/|B_1|$ is roughly
\begin{equation}
\label{l0}
l_0\sim\frac{\ln(16l_0^2/\epsilon)}{\ln(D/R_1)}\approx l_{00}+\frac{2\ln{l_{00}}}{\ln(D/R_1)},\quad l_{00}=\frac{\ln(16/\epsilon)}{\ln(D/R_1)}.
\end{equation}
Equation (\ref{l0}) is an upper bound. For $D/R_1=10$ and $\epsilon=0.001$ we have $l_0=6$.

Thus, the magnetic field outside two uniformly magnetized spheres with magnetic moments parallel to the axis passing through the centers of the spheres is given by~(\ref{axisymmetricB}), where the components of the magnetic field are given by (\ref{Beta}) and (\ref{Btheta}) with the constants being given by (\ref{alpha+}) and~(\ref{alpha-}). The validity of the calculated magnetic field is verified by direct substitution into the magnetostatic Maxwell equations (\ref{divBrotBzero}) and the boundary conditions~(\ref{continuousBn}).

\section{Discussion}

We see that the magnetic field $\mathbf{B}$ outside two uniformly magnetized spheres is not equal to the magnetic field $\mathbf{B}^\mathrm{d}$ of two dipoles with the magnetic moments
\begin{equation}
\label{magneticMoments}
\mathbf{m}_1=\frac{\mathbf{B}_1 R_1^3}2,\quad \mathbf{m}_2=\frac{\mathbf{B}_2 R_2^3}2
\end{equation}
placed in the centers of the respective spheres,
\begin{equation}
\label{Bdip}
\mathbf{B}\neq\mathbf{B}^\mathrm{d}=\mathbf{B}^\mathrm{d}_1+\mathbf{B}^\mathrm{d}_2,
\end{equation}
where
\begin{equation}
\label{Bdip12}
\mathbf{B}^\mathrm{d}_1=\frac{3\mathbf{m}_1\cdot\mathbf{r}_1\mathbf{r}_1}{r_1^5}-\frac{\mathbf{m}_1}{r_1^3},
\quad
\mathbf{B}^\mathrm{d}_2=\frac{3\mathbf{m}_2\cdot\mathbf{r}_2\mathbf{r}_2}{r_2^5}-\frac{\mathbf{m}_2}{r_2^3},
\end{equation}
$\mathbf{r}_1$ and $\mathbf{r}_2$ are the radius vectors directed to a given point from the centers of spheres 1 and~2, and $r_1$ and $r_2$ are their lengths.

The force acting on a sphere is equal to the integral of the Maxwell magnetic stress tensor \cite{Jackson1975}
\begin{equation}
\label{Tm}
\mathbf{T}_\mathrm{m}=\frac{\mathbf{B}\mathbf{B}}{4\pi}-\frac{B^2}{8\pi}\mathbf{I},
\end{equation}
where $\mathbf{B}\mathbf{B}=||B_i B_j||$ is the dyad and $\mathbf{I}=||\delta_{ij}||$ is the unit tensor, over any closed surface surrounding this sphere and not surrounding the other sphere, and the forces acting on spheres 1 and 2 are
\begin{equation}
\label{force12}
\mathbf{F}_1=\int_{S_1} \mathbf{T}_\mathrm{m}\cdot d\mathbf{S},\quad \mathbf{F}_2=\int_{S_2} \mathbf{T}_\mathrm{m}\cdot d\mathbf{S},
\end{equation}
where $S_1$ and $S_2$ are the surfaces surrounding the respective spheres and $d\mathbf{S}$ is the vector surface element, which is directed along the outward normal to the surrounding surface. It follows from the absence of magnetic monopoles $\nabla\cdot\mathbf{B}=0$, Amp\`{e}re's law $\nabla\times\mathbf{B}=4\pi\mathbf{j}/c$, and Gauss' theorem that the integrals entering (\ref{force12}) are equal to the integrals $(1/c)\int\mathbf{j}\times\mathbf{B}dV$ of Amp\`{e}re's forces acting on the respective spheres. The volume force density is zero inside the spheres and is singular at their surfaces, which corresponds to the finite surface forces generated by the azimuthal surface current determined from the discontinuity of the tangent magnetic field. Among many possibilities, we may take any sphere $\eta=const$ with $-\eta_1<\eta<0$ as $S_1$ and with $0<\eta<\eta_2$ as $S_2$ in~(\ref{force12}). Passing to the limit $\eta\rightarrow-0$ in the first case and $\eta\rightarrow+0$ in the second case, we are left in both cases with the integration over the same plane $\eta=0$, or $z=0$, but with the outward normal $\mathbf{e}_z$ in the first case and $-\mathbf{e}_z$ in the second case; therefore, the forces acting on the spheres are equal and oppositely directed, which is consistent with Newton's third law. Since the axial symmetry of the magnetic field implies $\int B_\theta\mathbf{e}_\theta \mathbf{B}\cdot d\mathbf{S}=0$, the forces are parallel to the $z$ axis and have the form
\begin{equation}
\label{forceFinal}
\mathbf{F}_1=-\mathbf{F}_2=F\mathbf{e}_z,
\end{equation}
where
\begin{equation}
\label{forceScalar}
F = \int\frac{(B_\eta|_{\eta=0})^2-(B_\theta|_{\eta=0})^2}{8\pi}d S
\end{equation}
and the integration is performed over the plane $\eta=0$, while the components of the magnetic field in this plane are
\begin{equation}
\label{BetaInPlane}
B_\eta|_{\eta=0} = \sqrt{\frac{1-\cos\theta}2}\sum_{l=1}^\infty \bigl[\mathrm{P}_l(\cos\theta) - \mathrm{P}_{l-1}(\cos\theta)\bigr] l (\alpha^+_l - \alpha^-_l),
\end{equation}
\begin{equation}
\label{BthetaInPlane}
B_\theta|_{\eta=0} = \sqrt{\frac{1-\cos\theta}2}\sum_{l=1}^\infty \bigl[\mathrm{P}^1_l(\cos\theta) - \mathrm{P}^1_{l-1}(\cos\theta)\bigr] (\alpha^+_l + \alpha^-_l).
\end{equation}

The difference between the fields $\mathbf{B}$ and $\mathbf{B}^\mathrm{d}$ implies that the interaction between the spheres in not equivalent to the interaction between the mentioned magnetic dipoles. The calculation of the interaction force via the Maxwell magnetic stress tensor, depending on the magnetic field only, also fits for the case of two dipoles with the only difference that the components of the magnetic field (\ref{Bdip}) should be substituted into~(\ref{forceScalar}). The forces acting on the dipoles become
\begin{equation}
\label{dipoleForce}
\mathbf{F}^\mathrm{d}_1 = -\mathbf{F}^\mathrm{d}_2 = \frac{6m_1 m_2}{D^4}\mathbf{e}_z,
\end{equation}
where $m_1=B_1 R_1^3/2$ and $m_2=B_2 R_2^3/2$ are the signed magnitudes of the magnetic moments $\mathbf{m}_1=m_1\mathbf{e}_z$ and $\mathbf{m}_2=m_2\mathbf{e}_z$.

The difference between the interactions of two spheres and two dipoles is brightly illustrated by the case of the interaction between a uniformly magnetized sphere (sphere 1 with $B_1\neq0$) and a nonmagnetized sphere (sphere 2 with $B_2=0$). The magnetic field near the nonmagnetized sphere is tangent to its surface because $B_\eta=0$ at the surface, and this situation is realized when the sphere is superconducting. Were these two spheres to interact as two point dipoles placed in their centers, the interaction force (\ref{dipoleForce}) would be equal to zero, $\mathbf{F}^\mathrm{d}_1 = \mathbf{F}^\mathrm{d}_2 = 0$, because $m_2=0$ corresponds to sphere~2. However, the real interaction force is nonzero, which becomes evident when the force is calculated by integration of $\mathbf{T}_\mathrm{m}$ over the sphere $\eta=const$ tightly embracing sphere~2, which corresponds to the limit $\eta\rightarrow\eta_2-0$:
\begin{equation}
\label{force0}
\mathbf{F}_1=-\mathbf{F}_2=\int p_\mathrm{m}|_{\eta=\eta_2-0} d\mathbf{S},
\end{equation}
where the integration is performed over the mentioned sphere $\eta=\eta_2-0$ and
\begin{equation}
\label{magneticPressure}
p_\mathrm{m}|_{\eta=\eta_2-0} = \frac{(B|_{\eta=\eta_2-0})^2}{8\pi} = \frac{(B_\theta|_{\eta=\eta_2-0})^2}{8\pi}
\end{equation}
is the magnetic pressure acting on sphere~2. Putting $B_2=0$ and $\eta=\eta_2$ in~(\ref{Btheta}), expanding the fractions in $\alpha^+_l$ and $\alpha^-_l$ into power series with respect to $e^{-(2l+1)(\eta_1 + \eta_2)}$ and $e^{-(2l-1)(\eta_1 + \eta_2)}$, and summing over~$l$, we eventually obtain the magnetic field inducing the pressure~$p_\mathrm{m}$,
\begin{equation}
\label{pressureField}
B_\theta |_{\eta=\eta_2-0} = -3\sqrt{2}\,B_1 (\cosh\eta_2-\cos\theta)^{3/2} \sin\theta \sum_{n=0}^\infty \frac{q_n^{3/2} (1-q_n^2)}{(1-2q_n \cos\theta+q_n^2)^{5/2}},
\end{equation}
where
\begin{equation}
\label{qn}
q_n = e^{-[2(n+1)\eta_1+(2n+1)\eta_2]}\leq e^{-2\eta_1-\eta_2}<1.
\end{equation}

Let us consider the magnetic field (\ref{pressureField}) as a function of the polar angle $\theta_2$ of the spherical coordinate system with the origin being the center of sphere~2 and the polar axis being the $z$ axis. We have from~(\ref{eEta})
\begin{equation}
\sin\theta=\frac{\sinh\eta_2\sin\theta_2}{\cosh\eta_2+\cos\theta_2},
\quad
\cos\theta=\frac{\cosh\eta_2\cos\theta_2+1}{\cosh\eta_2+\cos\theta_2},
\end{equation}
whence the magnetic field becomes
\begin{equation}
\label{pressureField2}
B_\theta |_{\eta=\eta_2-0}= -3\sqrt{2}\,B_1 \sinh^4\eta_2 \sin\theta_2 \,S,
\end{equation}
where the positive series
\begin{equation}
\label{Sseries}
S = \sum_{n=0}^\infty \frac{q_n^{3/2} (1-q_n^2)}{\bigl[(1+q_n^2)\cosh\eta_2 -2q_n + (1-2q_n\cosh\eta_2+q_n^2)\cos\theta_2\bigr]^{5/2}}>0
\end{equation}
is an increasing function of~$\theta_2$, as follows from the relation
\begin{equation}
\label{factorCos}
1-2q_n\cosh\eta_2+q_n^2=(1-q_n e^{\eta_2})(1-q_n e^{-\eta_2})>0,
\end{equation}
which in turn follows from (\ref{qn}) implying $q_n e^{\eta_2}<1$ and $q_n e^{-\eta_2}<1$.

\begin{figure*}
\centering
\includegraphics[width=8.5cm]{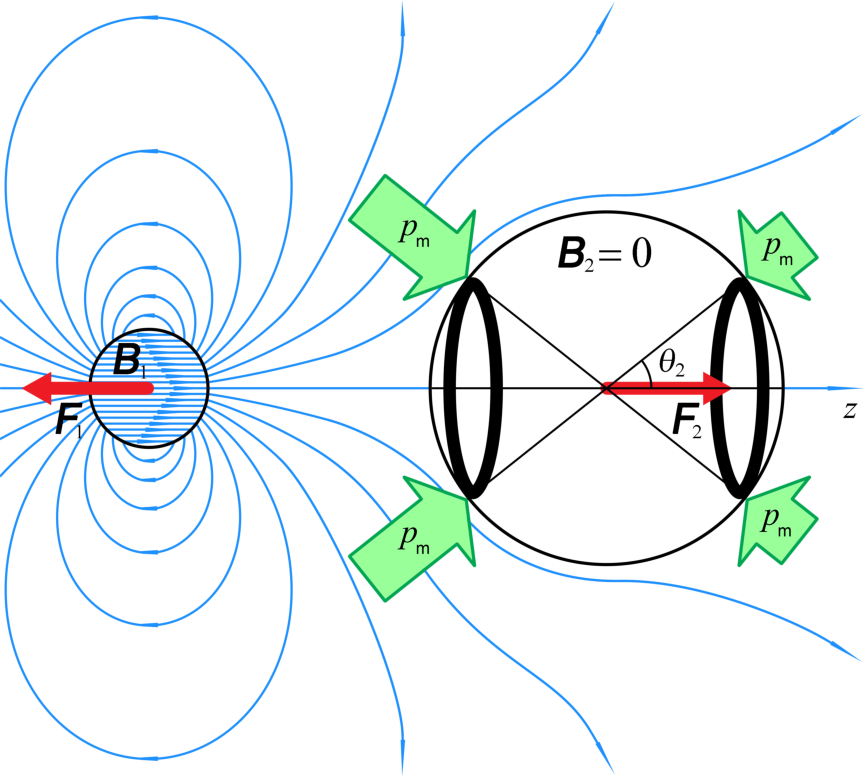}
\caption{Repulsion of a superconducting sphere from a uniformly magnetized sphere due to the magnetic pressure difference.}
\label{fig2}
\end{figure*}

The surface of the sphere represents a set of rings determined by the value of~$\theta_2$, and we may consider two rings symmetric with respect to the plane $z=z_2$ passing through the center of sphere 2 perpendicularly to the $z$ axis (Fig.~\ref{fig2}). It follows from (\ref{magneticPressure}) and~(\ref{pressureField2})--(\ref{factorCos}) that the difference of the absolute values of the magnetic fields and the difference of the magnetic pressures corresponding to the two symmetric rings with $\theta_2=\chi$ and $\theta_2=\pi-\chi$ for every $\chi$ from the range $0<\chi<\pi/2$ are nonzero and have the same sign,
\begin{equation}
\label{positiveDifferencesPressure}
B|^{\eta=\eta_2-0}_{\theta_2=\chi}<B|^{\eta=\eta_2-0}_{\theta_2=\pi-\chi},
\quad
p_\mathrm{m}|^{\eta=\eta_2-0}_{\theta_2=\chi}<p_\mathrm{m}|^{\eta=\eta_2-0}_{\theta_2=\pi-\chi}.
\end{equation}
Since $p_\mathrm{m}>0$ in the range $0<\theta_2<\pi$ and the projection of the outward normal $\mathbf{n}_2$ on $\mathbf{e}_z$ is positive for $\theta_2<\pi/2$ and negative for $\theta_2>\pi/2$, then due to the axial symmetry the integral $\int p_\mathrm{m}d\mathbf{S}$ over the hemisphere $\theta_2<\pi/2$ is nonzero and parallel to $\mathbf{e}_z$ and over the hemisphere $\theta_2>\pi/2$ is nonzero and antiparallel to~$\mathbf{e}_z$. We immediately conclude from (\ref{positiveDifferencesPressure}) that the sum of the magnetic pressure forces acting on the hemisphere $\theta_2>\pi/2$, which is closer to sphere~1, is larger than the sum of the forces acting on the farther hemisphere $\theta_2<\pi/2$, so that the total force is nonzero and repels sphere 2 along $\mathbf{e}_z$.

It follows from (\ref{axisymmetricB}), (\ref{Beta}), (\ref{Btheta}), (\ref{alpha+}), and (\ref{alpha-}) that in the general case of nonzero $B_1$ and~$B_2$ the magnetic field $\mathbf{B}$ outside two uniformly magnetized spheres is the sum of two magnetic fields, either of which depends on the internal magnetic field of only one of the two spheres,
\begin{equation}
\label{BB1B2}
\mathbf{B}=\mathbf{B}^0_1+\mathbf{B}^0_2,
\end{equation}
where $\mathbf{B}^0_1=\mathbf{B}|_{B_2=0}$ is proportional to $B_1$ and $\mathbf{B}^0_2=\mathbf{B}|_{B_1=0}$ is proportional to~$B_2$. However, $\mathbf{B}^0_1$ is not equal to the dipole magnetic field $\mathbf{B}^\mathrm{d}_1$, which sphere 1 generates when sphere 2 is absent, and represents the magnetic field into which the initially dipole magnetic field transforms when a superconducting sphere of radius $R_2$ is brought in the position of sphere~2; analogously, $\mathbf{B}^0_2$ is not equal to the dipole magnetic field $\mathbf{B}^\mathrm{d}_2$, which sphere 2 generates when sphere 1 is absent, and represents the magnetic field into which the initially dipole magnetic field transforms when a superconducting sphere of radius $R_1$ is brought in the position of sphere~1. The distortion of the dipole magnetic field occurs because the superconducting sphere in the ideal Meissner state displaces the surrounding magnetic field lines outward and does not allow them to penetrate inside. The same logic is applicable to two point magnetic dipoles, which can be considered as infinitesimal uniformly magnetized spheres such that $R_1\rightarrow+0$ and $R_2\rightarrow+0$ while $B_1=2m_1/R_1^3\rightarrow+\infty$ (or $-\infty$) for positive (or negative) $m_1$ and $B_2=2m_2/R_2^3\rightarrow+\infty$ (or $-\infty$) for positive (or negative) $m_2$ so that $m_1$ and $m_2$ remain finite and become the magnetic moments of the resulting point dipoles. In this case the total magnetic field becomes the sum of two dipole fields, $\mathbf{B}^0_1\rightarrow\mathbf{B}^\mathrm{d}_1$ and $\mathbf{B}^0_2\rightarrow\mathbf{B}^\mathrm{d}_2$, because an infinitesimal superconducting sphere is a point, which does not move the magnetic field lines and distort the dipole magnetic field. Thus, a finite size of at least one of the two uniformly magnetized spheres leads to the nondipole interaction between them, which is a reflection of the fact that a solution of the Laplace equation depends not only on the boundary conditions but also on the domain where the solution is being sought.

Passing to the limit $R_1\rightarrow+0$, corresponding to $\eta_1\rightarrow+\infty$, we can also consider the interaction between a point magnetic dipole (infinitesimal sphere 1 with magnetic moment~$\mathbf{m}_1$) and a superconducting sphere (nonmagnetized sphere~2). We obtain from (\ref{qn}) and (\ref{Sseries}) the asymptotics
\begin{equation}
\label{SseriesLim}
S \sim \frac{e^{-3\eta_1-3\eta_2/2}}{(\cosh\eta_2 + \cos\theta_2)^{5/2}},\quad \eta_1\rightarrow+\infty,
\end{equation}
which in combination with $e^{-\eta_1}\sim R_1/2z_1$ yields the magnetic field near sphere~2,
\begin{equation}
\label{BthetaDipole}
B_\theta|_{\eta=\eta_2-0} = -\frac{3}{2\sqrt{2}}\frac{m_1}{z_1^3}\frac{e^{-3\eta_2/2}\sinh^4\eta_2 \sin\theta_2}{(\cosh\eta_2 + \cos\theta_2)^{5/2}}.
\end{equation}
It remains to calculate the magnetic pressure (\ref{magneticPressure}), integrate according to~(\ref{force0}), and take into account the equality $\sinh\eta_2=a/R_2$ and the limiting relations $e^{\eta_2}=D/R_2$ and $a=z_1=(D^2-R_2^2)/2D$. We eventually obtain the interaction force between the magnetic dipole and the superconducting sphere,
\begin{equation}
\label{forceDipole}
\mathbf{F}_1=-\mathbf{F}_2= -\frac{6 m_1^2 D R_2^3}{(D^2-R_2^2)^4}\mathbf{e}_z,
\end{equation}
which agrees with the force calculated earlier using the method of images \cite{Lin2006,Palaniappan2009}.

The method of images allows us to see that the interactions of two spheres and two dipoles differ without complex formulas. To the dipole with magnetic moment $\mathbf{m}_1$ placed in the center of sphere~1, at point $(0,0,-z_1)$, there corresponds the image dipole with magnetic moment $\mathbf{m}_{11}=m_{11}\mathbf{e}_z=-(R_2/D)^3\mathbf{m}_1$ placed inside sphere 2 at the inverse point $(0,0,z_2-R_2^2/D)$ \cite{Levin1964,BeloozerovLevin1966,SezginerChew1990,LindellLehtola1992}. The magnetic field
\begin{equation}
\label{imageDipole}
\mathbf{B}^\mathrm{d}_{11}=\frac{3\mathbf{m}_{11}\cdot\mathbf{r}_{11}\mathbf{r}_{11}}{r_{11}^5}-\frac{\mathbf{m}_{11}}{r_{11}^3}
\end{equation}
corresponds to the image dipole, where $\mathbf{r}_{11}$ is the radius vector of length $r_{11}$ directed to a given point from the image dipole. We can directly verify that the total magnetic field of the initial dipole and its image is tangent to the surface of sphere~2,
\begin{equation}
\label{tangentField}
\mathbf{n}_2\cdot(\mathbf{B}^\mathrm{d}_1+\mathbf{B}^\mathrm{d}_{11})_{\eta=\eta_2-0}=0,
\end{equation}
so that
\begin{equation}
\label{B1B11}
(\mathbf{B}^\mathrm{d}_1+\mathbf{B}^\mathrm{d}_{11})_{\eta=\eta_2-0}=B_{\theta_2}\mathbf{e}_{\theta_2},
\end{equation}
where $\mathbf{e}_{\theta_2}=\mathbf{e}_\theta|_{\eta=\eta_2}$ is the unit tangent vector showing the direction of the increasing~$\theta_2$ and
\begin{equation}
\label{Btheta2}
B_{\theta_2}=-\frac{3m_1(D^2-R_2^2)\sin\theta_2}{(D^2+2DR_2\cos\theta_2+R_2^2)^{5/2}}.
\end{equation}

If we first consider the case $B_2=0$, introduction of the image dipole allows us to satisfy correct boundary conditions at the surface of sphere~2. Meanwhile, the image dipole generates a parasitic nonzero normal component of the magnetic field at the surface of sphere~1, violating the boundary conditions. To compensate the parasitic field, we should introduce another image dipole, the image of the first image dipole. It is placed inside sphere 1 at point $(0,0,-z_1+R_1^2/D_{11})$ and has magnetic moment $\mathbf{m}_{12}=-(R_1/D_{11})^3\mathbf{m}_{11}$, where $D_{11}=D-R_2^2/D$ is the distance from the center of sphere 1 to the first image dipole. The boundary conditions become satisfied at the surface of sphere 1 but become violated at the surface of sphere~2. To correct the violation, we should add the third image dipole, the image of the second image dipole, and so on. The resulting infinite series of successive images of the point dipole $\mathbf{m}_1$ gives the magnetic field satisfying the boundary conditions at the surfaces of both spheres when $B_2=0$. Adding an infinite series of successive images of the point dipole $\mathbf{m}_2$ gives the magnetic field satisfying the boundary conditions at the surfaces of both spheres in the general case of $B_1\neq0$ and $B_2\neq0$. Thus, the total magnetic field outside two uniformly magnetized spheres differs from the magnetic field of two point dipoles placed in their centers due to the existence of an infinite number of image dipoles with nonzero magnetic moments. The difference in the magnetic field results in the difference in the interaction.

When $B_2=0$, the resulting repulsion may be understood not only via the described action of magnetic pressure on the surface of superconducting sphere but also via the interaction of images. The first, third, and every odd image of the initial dipole $\mathbf{m}_1$ lies in superconducting sphere 2 and is antialigned with~$\mathbf{m}_1$, while the second, fourth, end every even image lies in magnetized sphere 1 together with $\mathbf{m}_1$ and is aligned with~$\mathbf{m}_1$. As a result, we have the interaction of two infinite sets of oppositely directed magnetic dipoles, each set containing dipoles of only one direction and lying in the respective sphere. Since by (\ref{dipoleForce}) two oppositely directed dipoles repel, so do these two infinite sets. The two ways of understanding are equivalent: in the case of the sole axially oriented point dipole, the magnetic field (\ref{BthetaDipole}) repelling the superconducting sphere coincides with the magnetic field (\ref{Btheta2}) of the dipole and its oppositely directed image, and the magnetic pressure force (\ref{forceDipole}) coincides with the dipole-image interaction force given by~(\ref{dipoleForce}),
\begin{equation}
\label{forceDipoleImage}
\mathbf{F}_1=-\mathbf{F}_2= \frac{6 m_1 m_{11}}{D_{11}^4}\mathbf{e}_z.
\end{equation}

In our definition, uniformly magnetized spheres are spheres with permanent internal magnetic fields satisfying~(\ref{B1B2parallelToZ}). Though we are not interested in the nature of these fields, nevertheless note a formal realization of this situation, two spheres of perfectly conducting medium. For the medium moving with velocity~$\mathbf{v}$, the internal magnetic field $\mathbf{B}_\mathrm{in}$ satisfies
\begin{equation}
\label{magneticTransport}
\frac{\partial\mathbf{B}_\mathrm{in}}{\partial t}=\nabla\times(\mathbf{v}\times\mathbf{B}_\mathrm{in}),
\end{equation}
a consequence of Faraday's law and Ohm's law for infinite conductivity \cite{Sobyanin2016}. If we assume that the two uniformly magnetized spheres are initially sufficiently distant from each other, when their mutual influence is negligible, but then close along the $z$ axis and occupy the final positions shown in Fig.~\ref{fig1}, so that their velocities are of the form $\mathbf{v}=v\mathbf{e}_z$, where $v=v(t)$ is a function of time only, different for each sphere, then it follows from (\ref{magneticTransport}) that
\begin{equation}
\label{dBdtZero}
\frac{d\mathbf{B}_\mathrm{in}}{d t}=\biggl(\frac{\partial}{\partial t}+\mathbf{v}\cdot\nabla\biggr)\mathbf{B}_\mathrm{in}=0.
\end{equation}
The zero substantial derivative means the constancy of internal magnetic fields during the motion of the spheres and reflects the well-known effect of magnetic freezing-in \cite{AlfvenFalthammar1963}.

Though the frozen-in internal magnetic field of one such sphere is not disturbed by the presence of the other sphere, this does not mean that the mutual influence of the spheres is absent. The electric currents flowing on the surfaces of the spheres and being determined from the jump in the transverse component of the magnetic field change as the spheres close, and so do the real magnetic moments associated with the respective spheres. Note also that even if sphere 2 is not internally magnetized, $B_2=0$, the existence of nonzero external magnetic field with negative $B_\theta$ (for positive $B_1$) implies the existence of surface currents $\mathbf{i}_2=i_2\mathbf{e}_\phi$ circulating counterclockwise when looking along the $z$ axis, $i_2=cB_\theta|_{\eta=\eta_2-0}/4\pi<0$; therefore, the real magnetic moment associated with sphere 2 (not the formal moment $\mathbf{m}_2=0$) is antialigned with the $z$ axis, $\mathbf{m}_2^\mathrm{real}=m_2^\mathrm{real}\mathbf{e}_z$ with negative $m_2^\mathrm{real}$, and so repels from the positive magnetic moment of sphere~1, as we have already seen while discussing the method of images.

\section{Conclusion}

We have examined the question whether the interaction between two uniformly magnetized spheres is equivalent to the interaction between two magnetic dipoles placed in the centers of the spheres, as is often stated. Using solely the Maxwell equations without any additional assumptions, we have calculated the magnetic field around the spheres whose internal magnetic fields are parallel to the axis passing through the centers. The components of the external magnetic field are represented in the bispherical coordinates via absolutely and uniformly convergent series of Legendre functions with exponential factors. This field fully determines the interaction force via the integral of the Maxwell stress tensor, and the difference between the fields of two spheres and two dipoles implies the difference between the interactions in these two non-equivalent cases. The non-equivalence reveals itself most clearly when one of the spheres is not magnetized: the dipole substitution leads to the zero interaction, while the exact result is the repulsion, which is consistent with the limiting result, the levitation of an axially oriented point dipole above a superconducting sphere.

\end{document}